# Distributed, Cross-Platform, and Regression Testing Architecture for Service-Oriented Architecture

Youssef Bassil

LACSC – Lebanese Association for Computational Sciences
Registered under No. 957, 2011, Beirut, Lebanon

Email: youssef.bassil@lacsc.org

**Abstract**– As per leading IT experts, today's large enterprises are going through business transformations. They are adopting service-based IT models such as SOA to develop their enterprise information systems and applications. In fact, SOA is an integration of loosely-coupled interoperable components, possibly built using heterogeneous software technologies and hardware platforms. As a result, traditional testing architectures are no more adequate for verifying and validating the quality of SOA systems and whether they are operating to specifications. This paper first discusses the various state-of-the-art methods for testing SOA applications, and then it proposes a novel automated, distributed, cross-platform, and regression testing architecture for SOA systems. The proposed testing architecture consists of several testing units which include test engine, test code generator, test case generator, test executer, and test monitor units. Experiments conducted showed that the proposed testing architecture managed to use parallel agents to test heterogeneous web services whose technologies were incompatible with the testing framework. As future work, testing non-functional aspects of SOA applications are to be investigated so as to allow the testing of such properties as performance, availability, and scalability.

## 1. Introduction

Many of today's enterprises are converting their information systems into new IT models based on e-services called Service-Oriented Architecture or SOA for short [1]. Fundamentally, SOA is the practice of designing and developing information systems using loosely-coupled interoperable software components [2]. SOA offers a number of benefits and advantages, such as flexibility, agility, reusability, scalability, maintainability, and interoperability [3]. However, adopting SOA comes with significant challenges, mostly related to the testing of SOA-based systems [4]. In fact, as SOA is an integration of several heterogeneous components, each built using different technologies and having incompatible interfaces, validating and verifying the operation of SOA can be viewed as a complex and challenging computing problem.

This paper presents a number of already existing approaches and techniques for testing SOA applications from different test levels including unit, integration, regression, distributed, and functional testing.

Furthermore, this paper proposes a new automated, distributed, cross-platform, and regression testing architecture for testing SOA applications and their web service components. It is made out of a test engine unit capable of conducting regression testing; a test code generator unit capable of generating client scripts for test execution; a test case generator unit capable of generating test conditions, variables, and data sequences; a test executer unit capable of applying test cases to web services; a test monitor unit capable of evaluating the testing results; and a database that stores valuable testing parameters throughout the testing process.

The proposed architecture has many benefits: It is distributed as it supports parallel testing of web services over multiple distributed server machines; it is cross-platform as it supports the testing of heterogeneous web services built using heterogeneous technologies; and it is capable of regression testing as it supports partial testing of sub-systems that have been recently changed or updated. All in all, the proposed architecture is meant to automate the testing of complex and heterogeneous SOA-based systems while achieving a good level of efficiency, performance, and quality.

## 2. Service-Oriented Architecture

Service-Oriented Architecture (SOA) is a model for system development based on loosely-integrated suite of services that can be used within multiple business domains [5]. Commonly, SOA is built using web service software components which are designed to support interoperable machine-to-machine interaction over a network. Predominantly, web services use SOAP (Simple Object Access Protocol), an XML-based protocol, to communicate over the HTTP protocol. Besides, they use WSDL (Web Service Description Language) to describe their internal functionalities and UDDI (Universal Description, Discovery, and Integration), a global registry and repository, to register and store their WSDLs [6]. Web services are governed by the producer-consumer/provider-requester model in which the provider owns the necessary equipment to host web services, and the requester connects to these web services and starts calling their exposed functions through method invocation mechanism. Several styles and types of web services exist, they include but not limited to SOAP, REST, .NET Remoting, RMI, RPC, and others. Figure 1 depicts the provider-requester model of a SOAP-based web service.

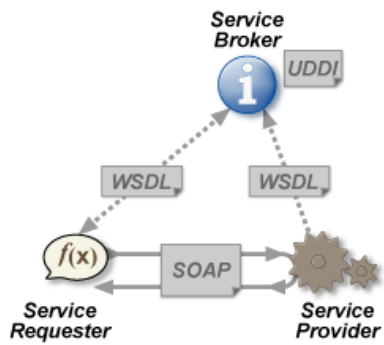

**Figure 1.** Provider-Requester Model of Web Services

## 3. Testing SOA

Software testing is an investigation carried out to determine whether a software is working correctly according to specifications [7]. The scope of software testing includes the validation and verification of the system's functional as well as non-functional properties. In that sense, the testing process can be defined as:

- Find if a software product is free of defects and is producing a correct output and;
- Find if a software product meets the customer's requirements as well as other technical requirements that guided its design and development.

As manual testing is a laborious and time consuming process, test automation has been employed thoroughly in many domains and fields. In essence, test automation uses software to perform, control, and monitor the execution of testing. It utilizes test cases which are set of input variables and their expected output that the test executer will apply to the software under test to determine whether it is working correctly according to specifications [8].

Since SOA-based systems are form of software, they should be tested too; however, since they are composed of a collection of fine-grained software components distributed over a network, they must be tested from a service-by-service viewpoint i.e. testing each web service of the SOA in isolation; from an end-to-end viewpoint i.e. testing the SOA as an aggregate of sub-systems; and from an interface-by-interface viewpoint i.e. testing the interoperability between the different web services of the SOA [9]. The different steps for testing SOA applications can be outlined as follows:

1. For a given SOA system under test, generate and execute a set of test requests.
2. Receive and evaluate the returned responses.
3. If the evaluation yields to a negative feedback, then the cause could be located in any of the web services that make up the SOA system under test:
   i. Repeat the above steps but for every service in isolation.
   ii. Find the malfunctioning web service, and refine it accordingly.
4. Repeat step 1 to step 3 until a positive feedback is obtained.

In practice, SOA can be tested using different test levels and techniques which can be summarized as follows [10]:

**Unit Testing:** It is the process of testing individual web services in isolation. The web service is disconnected from the SOA and tested separately in offline mode. Unit testing is usually conducted by developers to verify that the basic functionalities of web services are working correctly and according to specifications.

**Integration Testing:** It is the process of testing the SOA as a collection of web services that are working together in a group. It, in fact, focuses on testing web service interfaces to determine if communication and information sharing between them are working correctly and according to specifications.

**Regression Testing:** It is the process of re-testing an SOA that has been lately modified or updated to ensure that it does not fail due to the newly introduced repairs. Since each time a defect is fixed, there is a possibility that new errors get introduced, regression testing re-executes previously successful tests and checks whether previously working web services are still working correctly and according to specifications.

**Functional Testing:** It is the process of testing the basic functionalities of an SOA application. For example, testing if a web service that is exposing an addition function is able to add two numbers correctly and according to specifications.

**Non-Functional Testing:** It is process of testing the non-functional aspects of an SOA application which includes such properties as quality, performance, security, availability, interoperability, and other features already agreed on in the design specification stage of the SOA project.

## 4. SOA Testing Challenges

Testing SOA is somehow an intricate and a challenging computing problem, and that is due to several reasons, some of which are outlined below [11, 12]:

1. SOA are distributed in that they are composed of web service components dispersed over different hardware and operating system platforms; thus, testing must cover the different deployment configurations.
2. SOA are dynamic in that they implement adaptive behaviors such as adding new services, integrating new services, and removing old ones; consequently, performing an effective regression testing can be a challenging task.
3. SOA are complex in that they can be seen as a mesh of interacting services each having specific functionalities and capable of different operations; thus, designing test cases for test automation can be a complicated and a demanding task.
4. SOA are closed in that they are made out of closed services that run on the provider's side and clients have no control over their implementations; thus, preventing white-box testing methods that are essential to conduct exhaustive system validation.

5. SOA are remote in that their services are commonly located on the provider's server; and therefore, testing SOA can be costly, especially, if services are charged on a per-use basis. Moreover, service providers could suffer from denial-of-service (DoS) in case of massive testing.

6. SOA are heterogeneous in that their services deliver no standard interfaces for inter-communication as they are built using incompatible technologies, platforms, and programming languages; thus, it would be necessary to build multiple types of test engines each pertaining to a particular service platform.

# 5. Existing SOA Testing Approaches

This section reports the recent research achievements related to SOA testing including basic unit testing, distributed testing, testing by redundancy, integration testing, and regression testing.

## 5.1. Basic Unit Testing

A basic unit testing was proposed by [13]. The idea centers on testing individual web services using a test case generator and a test case executer. The proposed testing steps are as follows:

1. **Code Generation:** The necessary client code, also known as test script, is generated. Its purpose is to emulate a client consumer and to execute test cases.

2. **Test Case Generation:** A test case generation tool, the JCrasher, is used to generate test suites which are mainly composed of test cases.

3. **Test Execution:** The generated test cases are executed by invoking the various functions of the web services under test. Web services responses are then collected and validated against original system's specifications.

Figure 2 depicts the various modules of the basic unit testing approach.

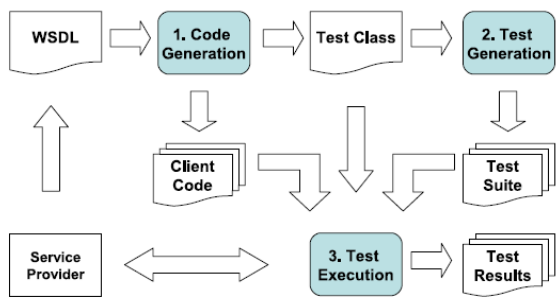

**Figure 2.** Typical Basic Unit Testing

Another basic unit testing technique was proposed by [14]. The approach uses a test case generator to generate test suites; a test engine executer to monitor the execution of test cases; and a log file to store the URL of the web service that has successfully passed the test. The proposed testing steps are as follows:

1. Connect to a particular UDDI registry, possibly located on the Internet, to search for a certain web service component to test. Once its WSDL is found, binding occurs between the web service, now called WSUT (Web Service under Test), and the testing framework.

2. The test case generator generates a test case that contains a series of function calls and data parameters. Afterwards, the test engine connects to the WSUT and executes the test cases by calling the functions of the WSUT through the SOAP protocol. The test engine then receives the response results from the WSUT and compares them with the expected results.

3. If both results match, then the WSUT is confirmed to pass the test, and its URL is saved into the log file; otherwise, the WSUT is confirmed to be defected and thus it is discarded.

4. The above steps are repeated for testing another web service.

Figure 3 depicts the inner-workings of this approach.

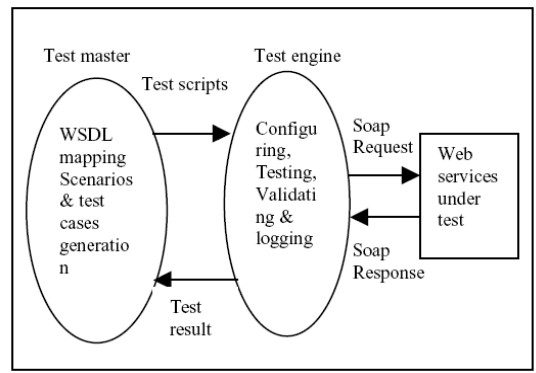

**Figure 3.** Another Typical Basic Unit Testing

### 5.1.1. Drawbacks

1. **Serial testing:** The generation and execution of test cases are done sequentially; parallelism or distribution of testing processes is not exploited.

2. **No support for regression testing:** In case of an update, all SOA components must be re-tested all over again.

3. **No support for integration testing:** All web services are tested in isolation; group testing is not exploited.

4. **Single-platform testing:** only SOAP-based web services can be tested; testing other types of web services such as REST or RMI is not exploited.

## 5.2. Distributed Testing

A distributed SOA testing approach was presented by [15] in which test cases are generated automatically based on the WSDL of the web service under test. The WSDL file is first parsed and transformed into a structured DOM tree. Then, test cases are generated and executed by a series of agents on distributed server machines. The approach employs multiple service brokers that can perform SOA testing simultaneously, each of which is equipped with a test case generator and a test execution controller. Below are the different units of this proposed distributed approach. Figure 4 shows the operation of these components.

1. **Test Case Generator:** It connects to the WSDL of the web service under test and automatically generates the necessary test cases which will be stored in a central database.
2. **Test Execution Controller:** It controls the execution of test cases in a distributed environment. Its job is to retrieve test cases from the test database, assign them to test agents, monitor test runs, and collect test results.
3. **Test Agents:** They are dispersed in a LAN or WAN area. An agent is a proxy that performs remote testing on target services with specific usage profiles and test data.
4. **Test analyzer:** It analyzes test results, evaluates the quality of services, and produces test reports.

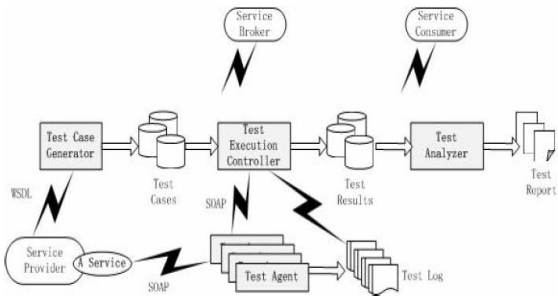

**Figure 4.** Distributed Testing for SOA

### 5.2.1. Drawbacks

1. **No support for regression testing:** In case of an update, all SOA components must be re-tested all over again.
2. **No support for integration testing:** All web services are tested in isolation; group testing is not exploited.
3. **Single-platform testing:** only SOAP-based web services can be tested; testing other types of web services such as REST or RMI is not exploited.

### 5.3. Testing by Redundancy

[16] proposed a collaborative redundancy-based verification and validation testing technique for SOA applications. In this approach, testing is conducted by evaluating multiple redundant web services at the same time. Then using a voter, the test engine compares the results of all web services under test. Only the web service, whose output is different from the other ones, is assumed to contain a fault. The advantage of this approach is that it does not require generating or implementing the client code. Evaluation is solely done by voting. Figure 5 shows the basic test architecture of this technique.

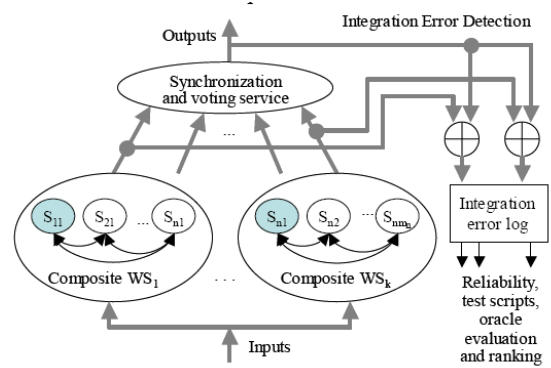

**Figure 5.** Testing By Redundancy

### 5.3.1. Drawbacks

Accuracy is dependent on the number of web services: The system will perform poorly if the number of web services is minimal; whereas, the precision of the voting system will increase as more web service are evaluated.

### 5.4. Integration Testing

[17] proposed an XML-based testing framework named Coyote for service integration testing in SOA. Coyote consists of two modules: a test master and a test engine. The test master allows testers to convert WSDL specifications into test scenarios and test cases, as well as performing non-functional analysis such as dependency, completeness, and consistency analysis. On the other hand, the test engine interacts with the web services under test, and provides tracing information. Integration testing is done during the development life cycle after the completion of the system specification phase. Every sub-system is tested using a stub proxy which houses all the required test suites and test scripts. The testing process goes top-down from the root SOA system to the leaf web service. This way, every child web service is verified whether it can communicate with its parent service. A top-down approach has two foremost advantages: It is easy to implement, and broken function calls and links can be discovered more efficiently. Figure 6 depicts the Coyote top-down testing approach.

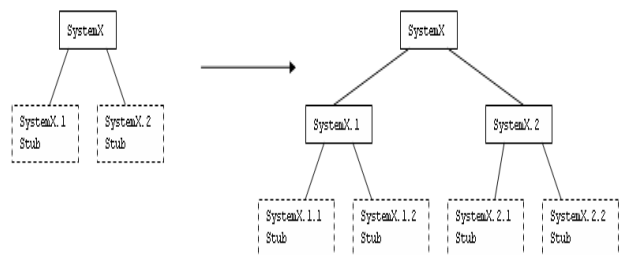

**Figure 6.** Coyote Integration Testing

### 5.4.1. Drawbacks

1. **Top-down approach:** Bugs are harder to be found and can interrupt the testing process.
2. **No support for regression testing:** In case of an update, all SOA components must be re-tested all over again.

3. **No support for integration testing:** All web services are tested in isolation; group testing is not exploited.

4. **Single-platform testing:** only single-protocol web services can be tested; testing multiple types of web services such as REST or RMI is not exploited.

## 5.5. Regression Testing

[18] proposed a regression testing approach for testing SOA applications. The role of regression testing is to uncover new software faults, called regressions, in existing parts of a system after changes have been made to them [19]. In this method, test cases are first generated, and then read by a test harness module which executes test cases over the various web services under test. The test harness module then collects the web service responses and stores them into a separate database to be later compared if they match the expected results. In case changes occur to the system, previously run tests are re-executed to check whether or not the behavior of the whole SOA system has changed. Figure 7 depicts the regression testing architecture for SOA systems.

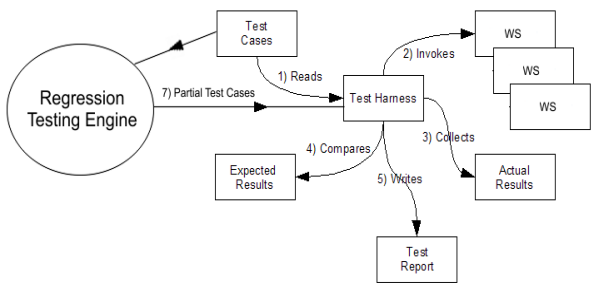

**Figure 7.** Regression Testing Architecture

### 5.5.1. Drawbacks

1. **Test suites complexity:** Test cases can be too large if changes come in too fast.
2. **Low performance:** Exhaustive test can increase the testing execution time and reserve a lot of resources.
3. **Single-platform testing:** only single-protocol web services can be tested; testing multiple types of web services such as REST or RMI is not exploited.

## 6. Proposed Testing Architecture

Based on the previous discussion of various SOA testing techniques, it is obvious that most of them can only support the testing of web services that are built using homogeneous technologies. However, it is no longer expected to test an SOA application that was designed using unified standards and protocols. Furthermore, all these previously discussed techniques are single-server single-machine systems, in that, testing multiple web services cannot be done in a parallel fashion but sequentially one after the other.

This paper proposes a new automated, distributed, cross-platform, and regression testing architecture for SOA systems. It supports the testing of multiple web services simultaneously using different instances of testing elements executed over distributed servers. In addition, it supports test planning and scheduling for multiple types of web services built using heterogeneous technologies, programming languages, and platforms. Finally, it supports regression testing by re-running previously executed test suites on the sub-systems that changes have been made to them. Below are the basic features and advantages of the proposed testing architecture:

- **Distributed:** It is capable of testing multiple web services concurrently on different machines, achieving better performance and higher throughput.
- **Cross-Platform:** It is capable of testing heterogeneous web services built using different platforms, different standards, and different programming languages.
- **Regression Testing:** It is capable of partial testing for system parts that have been changed recently, improving the quality assurance efficiency and reducing the time to re-validate SOA systems after changes have been made to them.

From a design standpoint, the proposed testing architecture consists of two parts: The part where the SOA application under test executes and the part where the testing framework executes. Figure 8 depicts these two parts together with their units and modules.

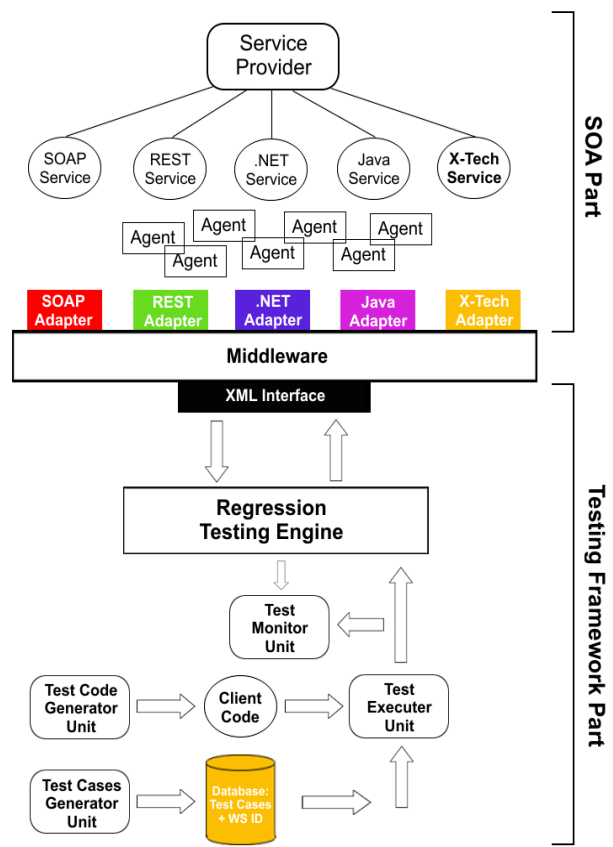

**Figure 8.** Parts of the Proposed Testing Architecture

## 6.1. The SOA Part

Essentially, the SOA part of the proposed testing architecture is majorly composed of web services under test, parallel agents, end-point adapters, and a central middleware.

The web services under test can be of any type and technology including SOAP, REST, .NET, Java or any X-technology; thus, providing a cross testing platform for SOA-based systems.

The parallel agents ensure a distributed testing by working as load balancers that distribute the test-load across multiple servers to achieve optimal resource utilization and to maximize throughput during the testing process. All test requests sent to web services under test are allocated to a free agent that, in turn, allocates them in a round robin fashion to any available back-end server machine to process the request. Agents constantly go on and off as test requests flow throughout the SOA.

Adapters are end-point connectors that bridge a test request with its destination service. They provide standardization and interoperability as they permit the interaction between the test engine and the different web services under test regardless of their underlying technologies and standards.

Adapters are supported by a central middleware that provides two interfaces: The first interface is from the SOA under test side which is mainly represented by the end-point adapters, and the second interface is from the testing framework side. The latter provides a unified XML interface to format test requests sent by the test engine to the web services under test. The test engine, through the middleware, sends XML-formatted test requests to web services under test regardless of their underlying technologies. The middleware then converts the received XML test requests into a format that is compatible with the addressed web service. As a result, the middleware provides a transparent communication between the test engine and the different web services of the SOA under test despite their incompatible technologies and platforms.

## 6.2. The Testing Framework Part

Basically, the testing framework is composed of several units each having a particular task to achieve and they are:

The test engine unit coordinates, controls, and manages the various testing units and their processes, and is capable of performing regression testing.

The test code generator unit generates test scripts and client code necessary to execute the test cases.

The test case generator unit generates all test scenarios, data suites, variables, and conditions necessary to create test requests and function calls and parameters for the web services under test.

The test executer unit executes the generated client code on the test cases, dispatches test requests, and collects testing results.

The test monitor unit evaluates and compares the results obtained from the test executer unit and the results obtained from the web services under test.

The database unit stores generated test cases along with the IDs of the web services under test, in addition to other miscellaneous testing parameters. In fact, the chief purpose of the database is to assist in the regression testing process. It permits the test engine unit to effectively retrieve and re-use the proper minimum set of previously stored test parameters and to execute them whenever a partial change has been made to the SOA application. That way, the test engine can check whether the SOA behavior has changed and whether previously fixed defects have recurred after fractional system updates.

## 7. Experiments & Results

As a proof of concept, a sample SOA application was tested using the proposed testing architecture. The target was to test a SOAP-based web service to prove the distributed testing capability of the parallel agents and the cross-platform testing capability of the middleware. The specifications for the conducted experimentation are below:

- **Web service under test:** A web service that contains an addition function to add two integer numbers.
- **Technology of the web service under test:** SOAP
- **ID of the web service under test:** 5
- **Test case:** Calling function $add(x, y)$ and sending integers 10 and 20 as test parameters.
- **Test code:** Implementation of the $add(x, y)$ function.

Below is the sequence of steps that were executed during the experimentation:

**Step 1:** The test code generator unit generated the client code, and stored it into the database together with the *ID=5* of the service under test.

**Step 2:** The test case generator unit generated the test case, namely the function call *add(10, 20)*, and stored it into the database together with the *ID=5* of the service under test.

**Step 3:** The test engine unit connected to the middleware and sent through it a test request to the SOAP-based web service under test in XML format. The test request represents a particular test case composed of a function call alongside with a set of data parameters, in this case *add(10, 20)*. The middleware first received the test request message and converted it from XML format into the protocol of the web service under test, in this case the SOAP protocol.

**Step 4:** The middleware routed the converted test request to the adapter that is compatible with the service under test, in this case, the SOAP adapter.

**Step 5:** The adapter then located a free agent to handle the test request. Once located, the free agent tried in sequence to locate the best machine on the network to process the request.

**Step 6:** The agent bound to the web service under test which executed the test request and returned back the integer *30* as the addition results of *10+20* in a SOAP message. The SOAP adapter received the response and routed it to the middleware. The middleware then converted the SOAP message into XML and forwarded it to the test monitor unit.

**Step 7:** The test executer unit executed the client code using the same test case, and reported the integer *30* as results.

**Step 8:** The test monitor unit then compared the results obtained from the web service response and the results obtained from the test executer unit. As both results matched, the executed test case was marked as successful in the database. Generally, if both results match, the corresponding test case is marked as successful in the database; otherwise, it is flagged as unsuccessful, and another test case is executed.

Below are two messages observed during the experimentation: The first is the XML test request sent by the test executer unit to the middleware, and the second is its equivalent SOAP message converted by the middleware and sent to the web service under test:

```
<request>
    <WS-ID>5</WS-ID>
    <function-to-call>add</function-to-call>
    <parameters>
        <param>10</param>
        <param>20</param>
    </ parameters >
    <timestamp>2/25/2012 05:22:17PM </timestamp>
</request>

<?xml version="1.0"?>
<soap:Envelope
    xmlns:soap="http://www.w3.org/2001/12/soap-envelope"
    soap:encodingStyle="http://www.w3.org/2001/12/soap-
encoding">
<soap:Body>
    <m:add>
        <m:x>10</m:x>
        <m:y>20</m:y>
    </m:add>
</soap:Body>
</soap:Envelope>
```

It is worth noting that in case of performing regression testing, only the test cases that were previously tested and marked as successful and whose ID matches the ID of web service that has been modified, are fetched from the database and re-used to re-evaluate this particular web service.

## 8. Conclusions & Future Work

This paper presented an extensive review for the various methodologies and research achievements related to SOA and web service testing. It, additionally, discussed the different challenges that SOA applications go through when testing their functional and non-functional properties. Furthermore, this paper proposed a novel architecture for testing SOA applications. It supports the following features: Distributed testing by employing parallel execution agents that can distribute the testing process over multiple machines; cross-platform testing by employing a central middleware that provides interoperability between the testing framework and the SOA application; and regression testing by employing a database that stores all successfully executed test cases so that they can be re-used to cover recent system modifications. Essentially, the testing framework comprises several testing units including test engine, test code generator, test case generator, test executer, and test monitor units. Experiments conducted showed that the proposed testing architecture managed to use parallel agents to allocate testing processes to distributed server machines and succeeded in exploiting its middleware to test heterogeneous web services whose technologies were incompatible with the testing framework.

As future work, testing non-functional aspects of SOA applications are to be investigated, bringing in a complete testing solution that can not only test SOA functional operations but also non-functional qualities such as performance, security, availability, and scalability.

## Acknowledgments

This research was funded by the Lebanese Association for Computational Sciences (LACSC), Beirut, Lebanon, under the "Simulation & Testing Research Project – STRP2012".